\documentclass[preprintnumbers,showpacs,showkeys,preprint,secnumarabic,amssymb,amsmath,amsfonts,aps, pre]{revtex4-1}
\usepackage{amsmath}
\usepackage{amsfonts}
\usepackage{enumitem}
\usepackage{yfonts}
\usepackage{mathrsfs}

\usepackage{amssymb}
\usepackage[english]{babel}
\usepackage[pdftex]{graphicx}
\usepackage{rotating}
\usepackage[pdftex,hypertexnames=true,linktocpage=true,breaklinks=true]{hyperref} 
\usepackage[hyphenbreaks]{breakurl}
\hypersetup{colorlinks=true,linkcolor=blue,anchorcolor=blue,citecolor=magenta,filecolor=blue,urlcolor=blue,bookmarksnumbered=false,pdfview=FitB}

\usepackage{esvect}

\newcommand{\affA}{Aix Marseille University, CNRS, CPT, Marseille, France}
\newcommand{\affB}{CNRS Centre de Physique Th\'eorique UMR7332,
13288 Marseille, France}

  \newcommand{\affG}{Quantum Biology Lab, Howard University, 2400 6th St NW, 16
17 Washington, DC 20059, USA}

\begin{document}
\title{Quantum Entanglement without nonlocal causation in (3,2)-dimensional spacetime}
 


\author{Marco Pettini}
\email{marco.pettini@cpt.univ-mrs.fr}
\affiliation{\affA}\affiliation{\affB}\affiliation{\affG}

\date{\today}

\begin{abstract}

This work aims at exploring whether the nonlocal correlations due to quantum entanglement could exist without nonlocal causation. This is done with the aid of a toy model to investigate whether the ability of two quantum entangled particles to ”correlate” their behaviors even at very large distances and in the absence of any physical connection can be seen as due to an exchange of information through an extra-temporal dimension. Since superluminal information exchange is forbidden in our (3,1) space-time, an extra-temporal dimension is needed to recover the physical picture of finite velocity information exchange between entangled entities. 
Assuming that the geometry of space-time of dimension (3,2) is described by a metric containing a warping factor, the confinement of the massive particles in the extra time dimension follows. Therefore, why we do not experience an infinitely large extra time dimension can be explained. 
The toy model proposed here is defined by borrowing Bohm-Bub’s proposal to describe the wavefunction collapse using nonlinear (non-unitary) dynamical equations and then elaborating this approach for an entangled system. \color{black}The model thus obtained aims to be a first step into unexplored territory, certainly a model that can be improved, but which already satisfies the purpose of giving the possibility to the hypotheses formulated above to be experimentally verified. The required experimental test has to resort to an unusual  experiment \color{black} which would otherwise be immediately dismissed as manifestly trivial.  The proposed experiment would consist of checking the possible violation of Bell’s inequality between two identical but independent systems under appropriate conditions. Beyond its theoretical interest, entanglement is a key topic in quantum computing and quantum technologies, so any attempt to gain a deeper understanding of it could be useful.

\end{abstract}

\maketitle

\section{Introduction}

Quantum nonlocality, related with non factorisability of the wavefunction of a composite system, leads to a seemingly paradoxical situation concerning the mutual action between particles or quanta, when a local and causal viewpoint is assumed.
The supposed incompleteness of quantum mechanics, motivated by the existence of the "spooky action at a distance" represented by the nonlocality paradox of the Einstein, Podolski, and Rosen (EPR) \textit{gedankenexperiment}, was refuted by Bohr's argument maintaining that a composite system must always be regarded as an indivisible totality which \textit{in principle} cannot be subdivided into independently existing units. 

Einstein's claim of incompleteness of quantum mechanics became susceptible of an experimental confirmation or refutation  after John Bell's proposal of a quantitative criterion \cite{bell} to test the completeness of the theory versus the need of the so-called hidden variables to recover a locality condition fulfilling Einstein’s causality. Under a suitable locality assumption, Bell’s theorem states that local hidden-variables theories are constrained by some given inequality, and some predictions of quantum mechanics can violate this inequality. A pioneering experimental implementation of a test based on Bell's criterion was put forward by Clauser, Horne, Shimony and Holt \cite{clauser,freedman} and led to a confirmation of quantum mechanics through the violation of Bell’s inequality.
However, having held fixed the orientation of the polarizers during the experimental run of this and other experiments made the results not conclusive because in the case of static experiments both the locality condition and the validity of Bell's inequality can be questioned. Bell thus insisted on the opportunity of performing experiments according to the Bohm-Aharonov proposal \cite{bohm0} of changing the settings of the instruments while the correlated particles were in flight. In so doing, the locality condition is entailed by Einstein's causality  forbidding faster-than-light interactions.

Then a strict and undeniable violation of Bell’s inequality was given by Alain Aspect's experiments \cite{aspect1,aspect2,aspect3} where each single channel polarizer was replaced by a fast switching device redirecting the incident light beam to a differently oriented polarizer. The detection events on opposite sides of the experimental apparatus were separated by a space-like interval, thus free of locality loopholes.
Further confirmations of the violation of Bell’s inequality were then obtained at increasingly large spatial separations \cite{nl1,nl2,nlz,nl3,nl4}.
Therefore, quantum nonlocality is undeniably a property of our physical world, but Bohr's viewpoint can sound tautologic and unsatisfactory compared to our  description of physical phenomena based on events located in space and evolving in time, so that the existence of correlations between   spatially separated events is necessarily attributed to some mutual influence that propagates in space with a finite velocity. This physical description is completely lost with quantum entanglement that implies instantaneous correlations between events separated by an arbitrarily large distance, meaning that mutual influences propagate at infinite velocity. Since this is a conundrum that goes against our intuitive perception and description of the physical world, some authors suggested to recover a realistic local model of quantum mechanics by resorting to superluminal communication between entangled objects at finite velocity $v>c$ and under appropriate conditions \cite{bohm,eberhard}.
A-priori this is not in contradiction with special Relativity because it can be shown that two observers performing experiments with entangled objects cannot use the corresponding quantum correlations to communicate at a faster-than-light speed \cite{bohm1}. This is referred to as no-signalling property of quantum correlations.
One of the first such proposal was formulated surmising that the superluminal influences at a distance are exerted via mechanisms involving an ether and effects propagating in that ether \cite{eberhard}. Said differently, the superluminal exchange of information leading to wave function collapse would be mediated by quantum tachyons, and to avoid causal paradoxes it has been shown that they have to propagate isotropically in a preferred reference frame with the same velocity in all directions \cite{liberati}. The compatibility between superluminal propagation and the fundamental principles of causality and relativity have been investigated \cite{liberati} and experimental attempts have been made to measure the velocity $v$ of quantum information in preferred reference frames \cite{scarani,salart,cocciaro} yielding lower bounds for $v>c$. However, in Ref.\cite{bancal} it was shown that "\textit{any possible explanation of quantum correlations in terms of influences propagating at any finite speed $v$}" with $c < v <\infty$ could violate the "\textit{impossibility of using non-local correlations for superluminal communication}". Hence to avoid conflict with Relativity theory $v$ cannot be finite and superluminal. The authors of this work conclude by affirming "\textit{[...] to keep no-signalling, [...] quantum non-locality must necessarily relate discontinuously parts of the universe that are arbitrarily distant. This gives further weight to the idea that quantum correlations somehow arise from outside spacetime, in the sense that no story in space and time can describe how they occur}". Although logically unescapable, this conclusion according to which a physical phenomenon originates outside of spacetime is disorienting and a source of conceptual discomfort, therefore, we can ask ourselves whether the statement "\textit{arise from outside [ordinary] spacetime}" can be replaced by "arise from inside an enlarged spacetime", in other words - after the "no go" for superluminal quantum communication at finite velocity - we can try to recover a "causal" description of quantum nonlocality by assuming the existence of an extra time dimension.
It is important to remark that we are not proposing a method that evades the requirement for a superluminal exchange of information, the projection to the ordinary (3+1) spacetime of the information exchange between two entangled objects through an extra temporal dimension of (3+2) spacetime  is superluminal but occurring at infinite velocity, thus respecting the no-signalling condition without any need for preferred frames. That is, the idea put forward in the present work leaves unchanged the nonlocality experienced in the ordinary (3+1) spacetime. 
\color{black}
In what follows there is no pretence of proposing something that could qualify as a new theory, 
but rather to put forward an unprecedented thought-provoking proposal to delve deeper into the mystery of quantum entanglement without contradicting the current explanation of quantum mechanics in any aspect.  From Dirac's axiomatic formulation of quantum mechanics in 1930 to nowadays quantum entanglement is considered a \textit{primitive} (i.e. "de facto") property of quantum formalism \cite{Qaxiom}.  Aside from the failed proposal of tachyon  communication, \color{black} no attempt has been made to explain entanglement from a more fundamental level. Therefore the proposal put forward in the present work is not an alternative explanation to something already existing. 
Actually,  by means of a simple (toy) model we move some first steps in an uncharted territory by exploring whether nonlocality can be given a "more familiar" interpretation by considering that while performing the measurement of an entangled state this appears as actually composed of independently existing units that exchange information via a hidden sub-quantum field ${\mathscr X}({\bf x}, t, \tau)$ that propagates at finite velocity through an extra dimension of temporal kind $\tau$. 
All this while respecting quantum contextuality, the property of quantum phenomenology proved by the Bell-Kochen-Specker theorem \cite{BKS,KS}.
This idea can be tested against an experiment that, if proved feasible and if it produced the hypothesized result, would make worthwhile and necessary to go beyond the toy model discussed below, \color{black} and would pave the way to the development of a topic of fundamental theoretical importance that nowadays may also have practical implications for cutting-edge research into quantum technologies. For example, the reply to questions like “\textit{is the degree of entanglement weakening at increasing distance?}”  is "no", according to a formalism describing as an indivisible system two entangled objects even if sitting at opposite borders of our galaxy \cite{nota}.  However, this formalism was developed to describe microscopic systems, and wondering whether it can be somewhat upgraded  is of conceptual interest and potentially practical relevance \cite{Qinternet}.
\begin{figure}[h!]
 \centering
 \includegraphics[scale=0.55,keepaspectratio=true,angle=-0]{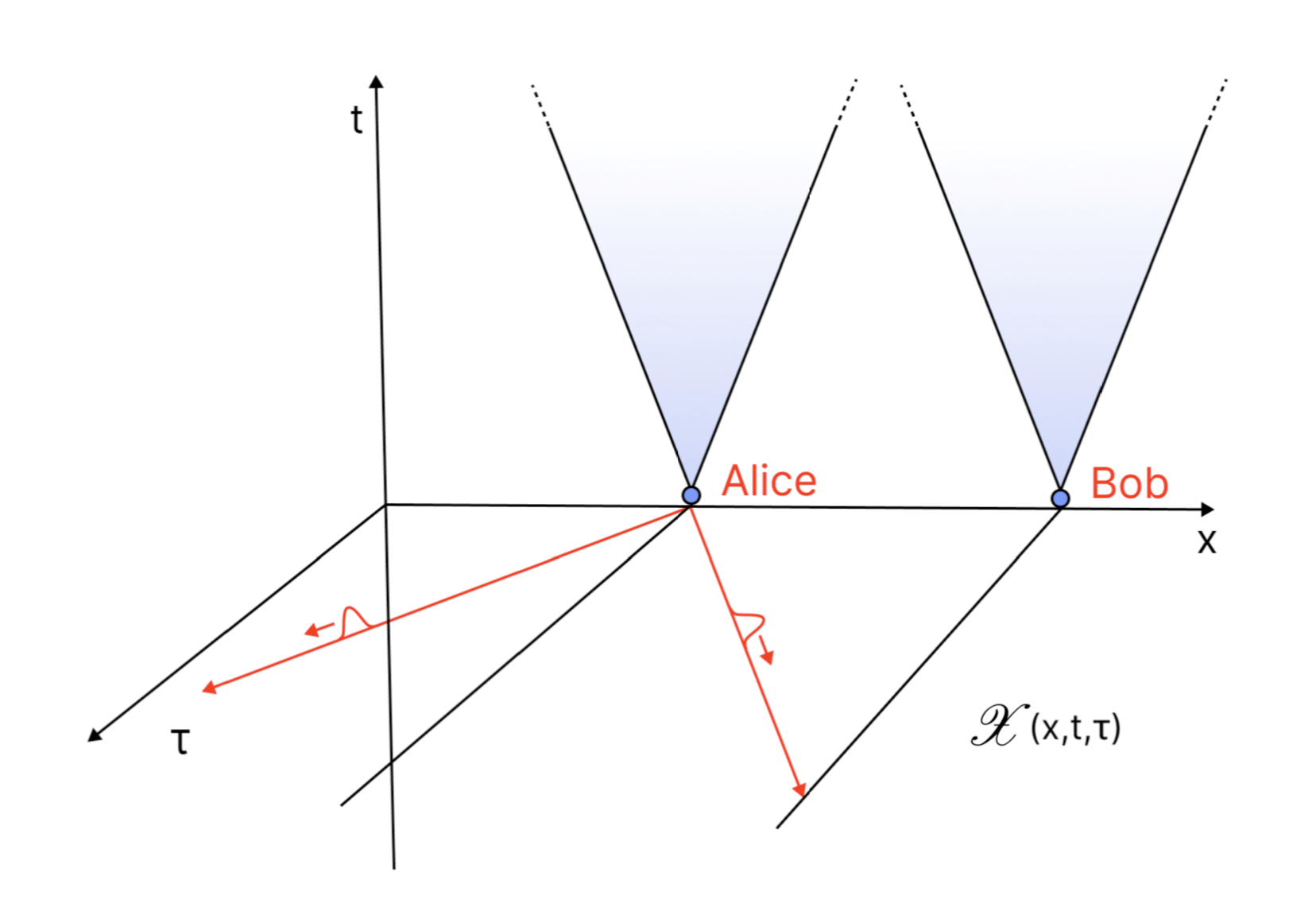}
 \caption{\sl{Schematic representation of a snapshot at measuring time. Alice and Bob are at a space-like distance (light cones are represented). A pulse-shaped signal of sub-quantum  field ${\mathscr X}({\bf x}, t,\tau)$ propagates through the extra time dimension $\tau$ carrying the information of the outcome of a measurement done by Alice.}}
\label{fig1}
\end{figure}
\subsection{A quick digression on extra time-dimensions}
The existence of extra dimensions, beyond the 3+1 with which we perceive the physical world, has entered theoretical physics with the formulation of the five-dimensional Kaluza-Klein theory (KKT), a classical unified field theory of gravitation and electromagnetism. With a few exotic exceptions \cite{rubakov,visser}, the fifth dimension - of space kind - in the KKT is compactifed under the so-called cylinder condition. 
The KKT is considered a precursor of string theory, where resorting to extra space dimensions is deemed natural and necessary \cite{rizzo}; also in this context the extra dimensions are curled up and microscopic. 
Now, what appears nonlocal in a (3,1) space-time could appear as such after projection from a higher dimensional space-time. This possibility has been suggested by considering extra dimensions of space kind surmising that \textit{``...while the usual fields only “live” in (3,1) dimensions, the collapse involves also other dimensions, eventually being induced by “some field” propagating also in these extra-dimensions''} \cite{genovese}. 
                                 
We might wonder why not considering the extra dimension of temporal kind. Actually, extra dimensions of time have been avoided because of several reasons. Among the others, in \cite{B1} it was claimed that with more than one time dimension, the partial differential equations for fields would be of ultrahyperbolic kind lacking the hyperbolicity property that enables observers to make predictions. 
However, it has been later proved that  the initial value problem for ultrahyperbolic equations, with data posed on an initial hypersurface of mixed
space and timelike signature, is well-posed \cite{B4}. Based on this work, the author of Ref.\cite{weinstein} proved that against to conventional beliefs, a well-posed initial value problem exists entailing deterministic, stable evolution for theories in multiple time dimensions. Moreover, the author puts forward the following intriguing idea: 
\textit{``quantum mechanics predicts nonlocal
entanglement between the properties of a given field at various locations in space. 
[...] The sort of constraint explored in this essay,
one arising from the presence of extra time dimensions, exhibits one sort of nonlocality, but there are
other sorts as well, [...] 
what I have called “nonlocality without nonlocality”, meaning nonlocal correlations without nonlocal causation.''}

On the other hand, any formulation of fundamental physics with multiple times is somehow non-trivial. In fact, as clearly summarized in Ref.\cite{B3}, naive attempts to add extra time dimensions to existing frameworks have led to two major kinds of discouraging and seemingly unavoidable difficulties: the appearance of ghosts and violations of causality. Ghosts are quantum states of systems occurring with negative probability. Causality violations are exemplified by the so called "grandfather paradox": two-dimensions of time by allowing time travels would make possible to kill one's ancestors before having being born, thus entailing an absurdity.
This notwithstanding, an extensive work (a survey of which can be found in Ref.\cite{bars4}) by Itzhak Bars and co-workers on a new gauge (symplectic) symmetry, called $Sp(2,{R})$, overcomes the mentioned problems and uniquely leads to the formalism of 2T-physics. In this two time theory the additional time dimension, being treated as a "gauge" and implied by the non-triviality of the new gauge principle, seems to some extent unphysical. 

It is worth mentioning that a new phase of matter \cite{nature2T} has recently been observed which seems to occupy two temporal dimensions. This new phase of matter was obtained in a quantum computer after emitting pulsed light on its qubits in a sequence following that of Fibonacci.
This experiment builds on earlier work that proposed the creation of something called a quasi-crystal in time \cite{dumitrescu}. Whether or not this result suggests the real existence of an additional time dimension is being discussed with a cautious attitude \cite{simons}.
In what follows we need to invoke the existence of a non-compact, arbitrarily large, extra time dimension. In this case, in order to avoid time loops, and thus to exorcise the specter of the “grandfather paradox”, we will  assume that only the field describing the sub-quantum level depends on two time dimensions since only the wavefunction collapse would be affected by the extra temporal dimension.
It is worth mentioning that in Ref.\cite{visser} in place of considering a compact extra space dimension, in an exotic version of Kaluza-Klein models an alternative has been explored in which the extra dimension is neither compact nor even finite, and particles are gravitationally trapped near a four-dimensional submanifold of the higher dimensional spacetime. In general, the warping of an extra dimension $\zeta$ added to $\{x^\mu\}$ is expressed as $ds^2= e^{-f(\zeta)}g_{\mu\nu} dx^\mu dx^\nu + d\zeta^2$ \cite{randall1,randall2}, where $f(\zeta)$ is a suitable function.

Similarly but with more details, in Ref.\cite{dvali} the authors investigate the phenomenology of extra time dimensions in presence of constraints that localize the standard model particles in the extra times, allowing them to move freely in our (3,1) space-time. This entails the breaking of the translation invariance in the extra time dimensions generating a Goldstone boson that propagates in all the space-time dimensions and is viewed as a tachyonic mode from our (3,1) space-time perspective.
The presence and the meaning of tachyonic modes in Kaluza-Klein models with extra time dimension(s) has been discussed from different viewpoints, see for instance Refs.\cite{yndurain,erdem}.

Summarizing, the possibility for a local dynamical theory in more dimensions than 3+1 to generate fundamentally non-local effects in lower dimensional space has been shown in several works where the extra dimensions (also additional times) play a key role (see also Ref. \cite{physrepGenovese}).

The paper is organized as follows. 
In Section \ref{nutshell} we sketch the Bohm-Bub theory, then in Section \ref{BBentangled} we propose an extension of this theory to simultaneous measurements of a Bell state. In Section \ref{experiment} we suggest that the fundamental idea proposed in the present work could be tested against an experiment. Section \ref{conclusions} contains some concluding remarks. Finally, in the Appendix it is sketchily shown how the formulation of the above mentioned exotic KKTs can be borrowed to define a warped metric of the (3,2)-dimensional spacetime yielding 
 the confinement of  massive particles in the extra time dimension, thus explaining why we do not experience an infinitely large extra time dimension. 
 In other words, the extra time dimension is assumed to be a  "hidden coordinate", that is, not observable even though physically existing. This prevents the appearance of ghosts and causality violations.
  
\section{Nonlinear equations for wavefunction collapse}
As is well known, two postulates of quantum mechanics are somewhat conflicting because on the one side the time evolution of the state vector of a given system is described by a linear, unitary operator and, on the other side, the result of the measure of an observable projects the system into the subspace relative to the eigenvalue/eigenstate of the measured observable according to the result obtained. Thus, measuring an observable entails the so-called collapse of the state vector (or wavefunction collapse), a  discontinuous breach of the unitary evolution of a quantum system, actually a nonlinear time evolution. This topic has given rise to several discussions and interpretations (see for instance Refs.\cite{bassi,RMP,dorje}), however, for our purpose we resort to the proposal described in the following section.
 
\subsection{The Bohm-Bub theory in a nutshell}\label{nutshell}
Our starting point is the Bohm-Bub non-unitary evolution model of the quantum state of a system during the measurement interaction with a macroscopic system.
 Bohm and Bub assume a non-unitary modification of the Schr\"odinger equation that reads  \cite{BB}
\begin{equation}
\label{schroedinger}
\frac{\partial\Psi({\bf x}, t)}{\partial t} = {\mathscr B}({\bf x}, t)  - \frac{i}{\hbar} \hat{H} \Psi({\bf x}, t)
\end{equation}
where ${\mathscr B}({\bf x}, t)$ has to take into account what happens during a measurement. In particular, during the interaction with a measuring apparatus
the term ${\mathscr B}({\bf x}, t)$ is assumed to be much larger than the standard unitary one yielding nonlinear dynamical equations that describe the wavefunction collapse. This term, with $\Psi({\bf x}, t) = \sum_i \psi_i(t) \phi_i({\bf x})$, after Ref.\cite{BB} and also after a more refined derivation \cite{tutsch}, is assumed to be
\begin{equation}\label{settoreB}
{\mathscr B}({\bf x}, t) =\gamma \sum_i \psi_i(t) \phi_i({\bf x}) \sum_j \vert\psi_{j}(t)\vert^2 (R_i - R_j)
\end{equation}
where $R_i=\vert\psi_i(t)\vert^2/\vert\xi_i(t)\vert^2$ and $\xi_i(t)$ are the components of the state vector of a hidden variable. Therefore, the Schr\"odinger equation becomes
\begin{equation}
\frac{d\psi_i(t)}{d t} = \gamma \psi_i(t) \sum_j \vert\psi_{j}(t)\vert^2 (R_i - R_j)  - \frac{i}{\hbar}\sum_j  {H}_{ij} \psi_j(t)
\end{equation}
and in the continuum case
\begin{equation}
\frac{\partial\psi({\bf x}, t)}{\partial t} = \gamma \psi({\bf x}, t) \int d{\bf y} \vert\psi({\bf y}, t)\vert^2 [R({\bf x}, t) - R({\bf y}, t)]  - \frac{i}{\hbar}\hat{H} \psi({\bf x}, t)\ .
\end{equation}
Applied to the special case of a dichotomic observable 
\begin{equation}\label{dichotomic}
\vert\psi\rangle = \psi_{+}(t)\vert S_+\rangle +  \psi_{-}(t)\vert S_-\rangle 
\end{equation}
after the introduction of a hidden state \cite{BB}
\[
\langle\xi\vert = \xi_{+}(t)\langle S_+\vert +  \xi_{-}(t)\langle S_-\vert
\]
whose components are randomly distributed hidden variables, the wavefunction collapse is described by the model equations 
 
\begin{equation}
\frac{d\psi_{+}(t)}{dt} = \gamma \left( R_{+} -  R_{-} \right)\psi_{+}( t) J_{-} - \frac{i}{\hbar} \left[ H_{++} \psi_{+}(t) + H_{+-} \psi_{-}(t)  \right]
\end{equation}
\begin{equation}
\frac{d\psi_{-}(t)}{dt} = \gamma \left( R_{-} - R_{+} \right)\psi_{-}(t) J_{+} - \frac{i}{\hbar} \left[ H_{--} \psi_{-}(t) + H_{-+} \psi_{+}(t)  \right]
\end{equation}
where $J_{\pm}=\vert\psi_{\pm}(t)\vert^2$, $R_{\pm}=\vert\psi_{\pm}(t)\vert^2/\vert\xi_{\pm}(t)\vert^2$, and the randomly distributed hidden variables are assumed constant under the assumption of an impulsive measurement.  During the measurement time 
 the interaction with the apparatus is assumed to be so large that the effects of the usual Schr\"odinger part - i.e. of the undisturbed system - can be neglected.
Hence, by multiplying the first equation by $\psi_{+}^\star$ and the second by  $\psi_{-}^\star$  one is left with the following nonlinear equations 
\begin{equation}\label{BB}
\frac{dJ_{+}}{dt} = 2 \gamma \left( R_{+} -  R_{-} \right)J_{+} J_{-} 
\end{equation}
\begin{equation}
\frac{dJ_{-}}{dt} = 2 \gamma \left( R_{-} - R_{+} \right)J_{-} J_{+} 
\end{equation}
whence $d(J_+ + J_-)/dt =0$ so that $\vert\psi\rangle$ remains normalized during measurement, and rewriting these equations as  
\begin{equation}
\frac{d\log J_{+}}{dt} = 2 \gamma \left( R_{+} -  R_{-} \right) J_{-} 
\end{equation}
\begin{equation}
\frac{d\log J_{-}}{dt} = 2 \gamma \left( R_{-} - R_{+} \right) J_{+} 
\end{equation}
where $\gamma$ is always positive,  if initially $R_{+} > R_{-}$ and $J_{-} \neq 0$ then $J_{+}$ increases and $J_{-}$ decreases until $J_{+}=1$ and
$J_{-}=0$ since $J_+ + J_- =1$; as a consequence the final state after the measurement of $\vert S\rangle$ is $\vert S_+\rangle$.
Conversely, if initially $R_{-} > R_{+}$ and $J_{+} \neq 0$ then the final state after the measurement of $\vert S\rangle$ is $\vert S_-\rangle$. The evolution of 
$\vert\psi\rangle$ during the measurement - and thus the final outcome of the measure - depends on the values that the hidden random variables 
$\xi_{+}(t)$ and $\xi_{-}(t)$ had immediately before the measurement, the outcome of which is therefore unpredictable.

\subsection{Extending Bohm-Bub's model to entangled particles }\label{BBentangled}

Let us now consider two particles, each one described by a dichotomic observable, that is 
\begin{eqnarray}\label{Phi1}
\vert\Phi^{(1)}\rangle &=& \psi^{(1)}_{+}(t)\vert\phi^{(1)}_{+}\rangle + \psi^{(1)}_{-}(t)\vert\phi^{(1)}_{-}\rangle\\
\vert\Phi^{(2)}\rangle &=& \psi^{(2)}_{+}(t)\vert\phi^{(2)}_{+}\rangle + \psi^{(2)}_{-}(t)\vert\phi^{(2)}_{-}\rangle \ ,
\end{eqnarray}
an entangled state of these particles is described by the Bell state
\begin{equation}\label{BellState}
\vert\Psi\rangle  =\frac{1}{\sqrt{2}}(\vert\phi^{(1)}_{+}\rangle\vert\phi^{(2)}_{-}\rangle - \vert\phi^{(1)}_{-}\rangle\vert\phi^{(2)}_{+}\rangle )\ ;
\end{equation}
as is well known, this means that a pair of entangled entities (particles, photons) must be considered a single non-separable physical object and it is impossible to assign local physical reality to each entity; this is a direct consequence of the formalism which implies that no physical theory explaining correlations between distant events by means of locality conditions can reproduce the quantum probabilities of the outcomes of experiments. This is certainly true in the four dimensional space-time, but we can hypothesize that what appears nonlocal in a 3+1 space-time can appear as such after projection from a higher dimensional space-time, as discussed in Section I.1. 

Before proceeding to extend the Bohm-Bub's model to entangled particles \cite{nota1}, a premise is necessary concerning the interaction of a microscopic system [as the one described in Eq.\eqref{BellState}] with the measuring apparatus (a macroscopic system). The latter can be thought of a many-body system described by a wave function factorized into a product of localized states of its constituent particles. When a microsystem combines with such a macroscopic system, a process of spontaneous localization occurs in the microsystem as it has been proposed in Ref.\cite{baracca}. A few years later
a more elaborated theory was proposed by Ghirardi, Rimini and Weber (GRW) to describe the spontaneous decoherence process of the quantum state describing a system with an arbitrary number of degrees of freedom $N$ \cite{GRW}, the decoherence time scale being proportional to 
$1/(N\lambda_{GRW})$, with $\lambda_{GRW}\sim 10^{-16} s^{-1}$, thus getting very short for a macroscopic system for which $N$ is large. 

 Let us now assume a time dependent $\gamma$ in ${\mathscr B}({\bf x}, t)$ of the form $\gamma(t) = N(t)\lambda_{GRW}$, where $N(t)$ is the number of particles with $N(t)=N_0 + N_{MA}\Theta(t - t_0)$ where $N_0$ is the number of particles of the quantum system, $N_{MA}$ is the number of particles of the measuring apparatus, 
 $\Theta(t - t_0)$ is a Heaviside step function, and $t_0$ is the time at which the measurement on the quantum system is performed. 


 As $N_{MA}$ is a macroscopic number, along the same line of thought proposed in Refs.\cite{baracca} and \cite{GRW} we can assume that at $t_0$ a complete  factorization of the composite state vector $\vert\Psi\rangle\vert\Psi_{MA}\rangle$ takes place, where $\vert\Psi\rangle$ is the microscopic system state vector and $\vert\Psi_{MA}\rangle$ is the  macroscopic state vector of the measuring device. In other words, when the microscopic system comes into contact with the measuring apparatus it becomes part of an overall system with a macroscopic number of degrees of freedom subject to the above mentioned decoherence mechanism. Therefore, since at the measuring apparatuses the particles are separated by a space-like interval, thus are distinguishable and non-interacting, the state vector \eqref{BellState}  factors into the product of the single particle states given above
 \begin{equation}
\vert\Psi\rangle = \vert\Phi^{(1)}\rangle \vert\Phi^{(2)}\rangle
\end{equation}
and - at the same instant of time $t_0$ -  in the Schr\"odinger equation ${\hat H}= {\hat H_1}\otimes{\hat H_2}$ splits as
 \begin{equation}
{\hat H} = {\hat H_1}\otimes{\hat{\mathbb I}_2} + {\hat{\mathbb I}_1}\otimes{\hat H_2}
\end{equation}
then the non-unitary evolution of the whole system (microscopic quantum system plus measuring apparatus) will read
\begin{equation}
\frac{d\psi^{(1)}_{+}(t)}{dt} = \gamma(t) \left( R^{(1)}_{+} - R^{(1)}_{-} + {R}^{(2)}_{-} - { R}^{(2)}_{+}\right)\psi^{(1)}_{+}(t) J^{(1)}_{-} - \frac{i}{\hbar} \left[ 
{H}_{1++} \psi^{(1)}_{+}(t) + H_{1+-} \psi^{(1)}_{-}(t) \right] \nonumber
\end{equation}
\begin{equation}
\frac{d\psi^{(1)}_{-}(t)}{dt} = \gamma(t) \left( R^{(1)}_{-} - R^{(1)}_{+}  + {R}^{(2)}_{+} - { R}^{(2)}_{-} \right)\psi^{(1)}_{-}( t) J^{(1)}_{+} - \frac{i}{\hbar} \left[ 
{H}_{1-+} \psi^{(1)}_{+}(t) + {H}_{1--} \psi^{(1)}_{-}(t) \right] \nonumber
\end{equation}
\begin{equation}
\frac{d\psi^{(2)}_+(t)}{dt} = \gamma(t) \left( R^{(2)}_{+} - R^{(2)}_{-} + { R}^{(1)}_{-} - { R}^{(1)}_{+}\right)\psi^{(2)}_{+}(t) J^{(2)}_{-} - \frac{i}{\hbar} \left[ {H}_{2++} \psi^{(2)}_{+}(t) + {H}_{2+-} \psi^{(2)}_{-}(t) \right] \nonumber
\end{equation}
\begin{equation}\label{collapse1}
\frac{d\psi^{(2)}_-(t)}{dt} = \gamma(t) \left( R^{(2)}_{-} - R^{(2)}_{+}  + { R}^{(1)}_{+} - { R}^{(1)}_{-} \right)\psi^{(2)}_{-}(t) J^{(2)}_{+} - \frac{i}{\hbar} \left[ {H}_{2-+} \psi^{(2)}_{+}(t) + {H}_{2--} \psi^{(2)}_{-}(t) \right]
\end{equation}
where 
\begin{equation}\label{erre2}
J^{(j)}_{\pm}=\vert\psi^{(j)}_{\pm}(t)\vert^2\ , \quad\quad\quad R^{(j)}_{\pm} = \frac{\vert\psi^{(j)}_{\pm}(t)\vert^2}{\vert\xi^{(j)}_\pm(t)  \vert^2}\ , \quad\quad\quad j=1, 2
\end{equation}
according to the initial sign of the terms in the first parenthesis of the r.h.s. of each one of the equations above, it is immediately evident that if the first particle collapses to the state $\vert +\rangle$ the second one collapses to $\vert -\rangle$ and viceversa. In analogy with the original Bohm-Bub's model, the $\xi^{(j)}_\pm(t) $ are assumed to be hidden random variables, almost everywhere non vanishing, and constant during measurement. The non-locality is here expressed by the fact that even if the two particles are separated by a space-like interval, the values of the quantities $R^{(j)}$ have an instantaneous mutual influence on the respective wavefunction collapses.
\color{black} Hereafter we introduce the following working hypothesis which is the theoretical core that is being considered. 
We assume that the correlated wavefunctions-collapses of the two particles - that now have their own individuality -  are driven by \color{black} an exchange of information mediated by a hidden sub-quantum field ${\mathscr X}({\bf x}, t,\tau)$ which propagates through an extra temporal  dimension described by the variable $\tau$. \color{black} This hypothesis necessarily requires the extra time dimension because a superluminal exchange of information is forbidden in our standard space-time. \color{black} We are replacing the hidden variables $\xi^{(j)}_\pm (t)$ with a completely different physical entity, that is, a real hidden physical vector field ${\mathscr X}({\bf x}, t,\tau)$, of components 
${\mathscr X}_\pm({\bf x}, t,\tau)$, almost nowhere vanishing (that is with the possible exception of a zero measure subset of space), a condition that could be ensured for example by the existence of a stochastic background component of the field. 
\color{black} Let us remark that the variables $\xi_i(t)$ of the original Bohm-Bub model, as well as the field ${\mathscr X}({\bf x}, t,\tau)$ introduced here, play a role only during the nonlinear, non-unitary dynamics of the state reduction, thus they do not belong to the "family" of hidden variables theories where quantum mechanics is assumed to be a statistical approximation of some unknown deterministic theory where the values of the observables are already defined and fixed by unknown, hidden variables.
 \color{black}

We tentatively assume that the dynamical evolution of the  field  ${\mathscr X}({\bf x}, t,\tau)$  is described by a wave equation
\begin{equation}
\nabla^2  {\mathscr X}({\bf x}, t,\tau) - \frac{1}{w^2}\frac{\partial^2{\mathscr X}({\bf x}, t,\tau)}{\partial\tau^2}\ +\  F[\psi_{\pm}^{(j)}(t),{\mathscr X}({\bf x}, t,\tau)] = 0
\label{evolEqForXi}
\end{equation}
accounting for information propagation at speed $w$ by means of ${\mathscr X}({\bf x}, t,\tau)$ through the dimension $\tau$, and where the inhomogeneous source term $F[\psi_{\pm}^{(j)}(t),{\mathscr X}({\bf x}, t,\tau)]$ should be chosen so that equation \eqref{evolEqForXi} fulfils the following requirements: \textit{i)} it describes the propagation of the field ${\mathscr X}({\bf x}, t,\tau)$ in the form of a non-dispersive impulse carrying the information of the outcome of a measure performed at some spatial location; \textit{ii)} this information is isotropically propagated in space; \textit{iii)} the information impulse carried by the fields ${\mathscr X}_\pm({\bf x}, t,\tau)$ is not attenuated with distance, and this assumption is required to comply with standard quantum non-locality which is independent of the distance between entangled entities. 
So we update equation \eqref{evolEqForXi} as
\begin{equation}\label{eqperxi}
  \nabla^2  {\mathscr X}_\pm(r, t,\tau) - \frac{1}{w^2}\frac{\partial^2{\mathscr X}_\pm(r, t,\tau)}{\partial\tau^2}\ +\  \sin[{\mathscr X}_\pm(r, t,\tau)] + {\widetilde{F}}[{\bf x}, \psi_{\pm}^{(j)}(t), \tau ; t_0, \tau_0] = 0
\end{equation}
that is a  wave equation where a tentative nonlinearity is introduced to account for the propagation of a non dispersive impulse, actually a sine-Gordon soliton, in a radial direction $r$ (under spherical symmetry), stemming from a given point ${\bf x}$, and carrying the information coded by the source term ${\widetilde{F}}[{\bf x}, \psi_{\pm}^{(j)}(t), \tau ; t_0, \tau_0]$, where $t_0$ and $\tau_0$ stand for the initial times. 
In analogy with the standard assumption \cite{belinfante} for the evolution of $\xi (t)$, that is, $\dot\xi (t)=0$ during an impulsive  measurement, no term containing the derivative of ${\mathscr X}({\bf x}, t,\tau)$ with respect to our usual time is considered in the equation above.
At the present state of affairs, we are tackling a toy model because we lack physical input to make more definite formulations, in particular to assign the analytical form of the function ${\widetilde{F}}[{\bf x}, \psi_{\pm}^{(j)}(t), \tau ; t_0, \tau_0] $, nevertheless, for the moment being it is sufficient to formally enter this term in Eq.\eqref{eqperxi} to represent the source of the information carried by  ${\mathscr X}({\bf x}, t,\tau)$.
In fact, already at the present stage an experiment seems conceivable (see the next section)  to test whether or not the surmised scenario corresponds to the  physical reality, in case of a positive experimental outcome one could make more motivated assumptions. 

Let us now rewrite Eq.\eqref{schroedinger} as 
\begin{equation}
\label{schroed}
\frac{\partial\Psi({\bf x}, t)}{\partial t} = {\widetilde{\mathscr B}}({\bf x}, t, {\overline\tau})  - \frac{i}{\hbar} \hat{H} \Psi({\bf x}, t)
\end{equation}
and rewrite the non-unitary evolution term \eqref{settoreB} as
\begin{equation}\label{settoreBtilde}
{\widetilde{\mathscr B}}({\bf x}, t, {\overline\tau}) =\gamma \sum_i \psi_i(t) \phi_i({\bf x}) \sum_j \vert\psi_{j}(t)\vert^2 ({\widetilde R}_i - {\widetilde R}_j)
\end{equation}
where the hypothetical information-field ${\mathscr X}({\bf x}, t, \tau)$ enters the terms ${\widetilde R}$ by replacing the hidden variables $\xi_i(t)$.
We remark that the term ${\widetilde{\mathscr B}}({\bf x}, t, {\overline\tau})$ is not  continuously dependent on $\tau$ but it depends just on point values of ${\mathscr X}({\bf x}, t, \tau)$ at some
 fixed extra time ${\overline\tau}= \tau_0 + \Delta\tau$, as is specified in the following, therefore this explains why in 
Eq.\eqref{schroed} $\Psi({\bf x}, t)$ is not replaced by $\Psi({\bf x}, t, \tau)$, in other words this means that we do not need a two-time extension of standard quantum mechanics. 

Then, with the notations of Eq.\eqref{settoreBtilde}, the collapse equations \eqref{collapse1} are replaced by  \color{black}
\begin{equation}\label{collapse1a}
\left.\frac{d\psi^{(1)}_{+}(t)}{dt}\right\vert_{P_1} = \gamma(t) \left( R^{(1)}_{+} - R^{(1)}_{-} + {\widetilde R}^{(2)}_{-} - {\widetilde  R}^{(2)}_{+}\right)\psi^{(1)}_{+}(t) J^{(1)}_{-} - \frac{i}{\hbar} \left[ {H}_{1++} \psi^{(1)}_{+}(t) + {H}_{1+-} \psi^{(1)}_{-}(t) \right] 
\end{equation}
\begin{equation}\label{collapse1b}
\left.\frac{d\psi^{(1)}_{-}(t)}{dt}\right\vert_{P_1} = \gamma(t) \left( R^{(1)}_{-} - R^{(1)}_{+}  + {\widetilde R}^{(2)}_{+} - {\widetilde  R}^{(2)}_{-}  \right)\psi^{(1)}_{-}(t) J^{(1)}_{+} - \frac{i}{\hbar}\left[ {H}_{1-+} \psi^{(1)}_{+}(t) + {H}_{1--} \psi^{(1)}_{-}(t) \right] 
\end{equation}
\begin{equation}\label{collapse2a}
\left.\frac{d\psi^{(2)}_+(t)}{dt}\right\vert_{P_2} = \gamma(t) \left( R^{(2)}_{+} - R^{(2)}_{-} + {\widetilde  R}^{(1)}_{-} - {\widetilde  R}^{(1)}_{+}\right)\psi^{(2)}_{+}(t) J^{(2)}_{-} - \frac{i}{\hbar} \left[ {H}_{2++} \psi^{(2)}_{+}(t) + {H}_{2+-} \psi^{(2)}_{-}(t) \right] 
\end{equation}
\begin{equation}\label{collapse2b}
\left.\frac{d\psi^{(2)}_-(t)}{dt}\right\vert_{P_2} = \gamma(t) \left( R^{(2)}_{-} - R^{(2)}_{+}  + {\widetilde  R}^{(1)}_{+} - {\widetilde  R}^{(1)}_{-}  \right) \psi^{(2)}_{-}(t) J^{(2)}_{+} - \frac{i}{\hbar} \left[ {H}_{2-+} \psi^{(2)}_{+}(t) + {H}_{2--} \psi^{(2)}_{-}(t) \right]
\end{equation}
where, with an abuse of notation, it is evidenced that the first two equations describe the measurement process at the space-time coordinates $P_1 =({\bf x}_1,t_0,\tau_0+\Delta\tau)$, and the other two equations describe the measurement process at the space-time coordinates $P_2 =({\bf x}_2,t_0,\tau_0+\Delta\tau)$, that is, one of the polarizers is located at ${\bf x}_1$ and the other polarizer is located at ${\bf x}_2$; both measurements are simultaneous, performed at $t_0$ and at $\tau_0+\Delta\tau$ where $\Delta\tau = \vert{\bf x}_2  - {\bf x}_1\vert/\vert{\bf w}\vert$ with ${\bf w}$ the velocity vector of information transfer between the particles through the extra time dimension, that is, the extra time at which a mutual exchange of information takes place.
In order to describe this exchange of information between the two subsystems - operated via the sub-quantum field ${\mathscr X}({\bf x}, t,\tau)$ propagating through the extra time dimension $\tau$ - and in order to keep the essential of the Bohm-Bub model the variables ${\widetilde R}^{(1)}_{\pm}$ and ${\widetilde R}^{(2)}_{\pm}$ in the equations above are assumed
to be of the form
\begin{equation}
{\widetilde R}^{(1)}_{\pm} = \frac{\vert\psi^{(1)}_{\pm}(t)\vert^2}{\vert{\mathscr X}_\pm({\bf x}_2 - {\bf w}\Delta\tau , t_0, \tau_0 + \Delta\tau )  \vert^2}\ , \quad\quad\quad
{\widetilde R}^{(2)}_{\pm} = \frac{\vert\psi^{(2)}_{\pm}(t)\vert^2}{\vert{\mathscr X}_\pm({\bf x}_1 - {\bf w}\Delta\tau , t_0, \tau_0 + \Delta\tau )  \vert^2}
\end{equation}
and
\begin{equation}
R^{(1)}_{\pm} = \frac{\vert\psi^{(1)}_{\pm}(t)\vert^2}{\vert{\mathscr X}_\pm({\bf x}_1, t_0, \tau_0 + \Delta\tau )  \vert^2}\ , \quad\quad\quad
R^{(2)}_{\pm} = \frac{\vert\psi^{(2)}_{\pm}(t)\vert^2}{\vert{\mathscr X}_\pm({\bf x}_2, t_0, \tau_0 + \Delta\tau )  \vert^2}
\end{equation}
with the assumption $\gamma(t) = [N_0 + N_{MA}\Theta(t - t_0)] \lambda_{GRW}$, where $N_0$ is the number of particles of the quantum system. 

 Notice that the field ${\mathscr X}({\bf x}, t, \tau)$ is assumed to propagate at finite (unknown) velocity $w$ according to equation (20) through the extra time $\tau$ only, and it is assumed to propagate at infinite velocity in our standard (3+1) spacetime. In fact, the denominators 
$\vert{\mathscr X}_\pm({\bf x}_{1,2} - {\bf w}\Delta\tau , t_0, \tau_0 + \Delta\tau )\vert^2$ - entering the above defined functions ${\widetilde R}^{(1)}_{\pm}$ and ${\widetilde R}^{(2)}_{\pm}$ - account for the propagation of ${\mathscr X}({\bf x}, t, \tau)$ only in the extra time dimension because the usual time enters solely with $t_0$, the instant of  measurement, thus implicitly assuming nonlocal instantaneous correlation in $t$ with the absence of a finite propagation speed.
This is an important point to ensure the no-signalling condition discussed in the Introduction.

\color{black} 
It is important to remark that the modifications of the Schr\"odinger  equation in \eqref{schroedinger} and \eqref{schroed} is without any consequence on the standard unitary evolution of a quantum system - including the Bell state \eqref{BellState} - until $\gamma(t) = const = N_0\lambda_{GRW}$ takes a large value because of the interaction with the measuring apparatus. In other words, neither the field ${\mathscr X}({\bf x}, t,\tau)$ nor the extra temporal dimension have any consequence on the standard unitary evolution until measurement.

In this latter case, under the hypothesis of the wavefunction factorization induced by the GRW mechanism,  the evolution of the quantum system during the measurement process
is described by equations \eqref{collapse1a}-\eqref{collapse2b} and as in the Bohm-Bub theory it is assumed that the unperturbed quantum evolution term is negligible in comparison with the interaction term between the quantum system and the measuring apparatus. Hence, the time evolution of the wavefunctions of the system is described only by the nonlinear parts of the equations above, that is, after multiplications by the complex conjugate of the wavefunction of each equation, by the following system of equations

\begin{eqnarray}\label{BBextended}
\frac{dJ^{(1)}_{+}}{dt} &=&2 \gamma (t)\left( R^{(1)}_{+} - R^{(1)}_{-} + {\widetilde R}^{(2)}_{-} - {\widetilde R}^{(2)}_{+}\right)J^{(1)}_{+} J^{(1)}_{-} \nonumber \\
\frac{dJ^{(1)}_{-}}{dt} &=& 2\gamma (t)\left( R^{(1)}_{-} - R^{(1)}_{+}  + {\widetilde R}^{(2)}_{+} - {\widetilde R}^{(2)}_{-}  \right)J^{(1)}_{-} J^{(1)}_{+} \\
\frac{dJ^{(2)}_+}{dt} &=& 2\gamma (t)\left( R^{(2)}_{+} - R^{(2)}_{-} + {\widetilde R}^{(1)}_{-} - {\widetilde R}^{(1)}_{+}\right)J^{(2)}_{+} J^{(2)}_{-}  \nonumber\\
\frac{dJ^{(2)}_-}{dt} &=& 2\gamma (t)\left( R^{(2)}_{-} - R^{(2)}_{+}  + {\widetilde R}^{(1)}_{+} - {\widetilde R}^{(1)}_{-}  \right)J^{(2)}_{-} J^{(2)}_{+} \nonumber 
\end{eqnarray}
which evidently preserve the normalization of wavefunctions, that is, $d(J^{(1)}_{+} + J^{(1)}_{-})/dt =0$ and $d(J^{(2)}_{+} + J^{(2)}_{-})/dt =0$.
Equations \eqref{BBextended} are rewritten as 
\begin{eqnarray}\label{BBlogextended}
\frac{d\log J^{(1)}_+}{dt} &=&2 \gamma (t)\left( R^{(1)}_{+} - R^{(1)}_{-} +  {\widetilde R}^{(2)}_{-} - {\widetilde R}^{(2)}_{+}\right) J^{(1)}_{-} \nonumber \\
\frac{d\log J^{(1)}_- }{dt} &=& 2\gamma (t)\left( R^{(1)}_{-} - R^{(1)}_{+}  + {\widetilde R}^{(2)}_{+} - {\widetilde R}^{(2)}_{-}  \right) J^{(1)}_{+} \nonumber \\
\frac{d\log J^{(2)}_+}{dt} &=& 2\gamma (t)\left( R^{(2)}_{+} - R^{(2)}_{-} +  {\widetilde R}^{(1)}_{-} - {\widetilde R}^{(1)}_{+} \right) J^{(2)}_{-}  \\
\frac{d\log J^{(2)}_- }{dt} &=& 2\gamma (t)\left( R^{(2)}_{-} - R^{(2)}_{+}  + {\widetilde R}^{(1)}_{+} - {\widetilde R}^{(1)}_{-}  \right)J^{(2)}_{+} \nonumber 
\end{eqnarray}
if $J^{(1)}_{-}\neq 0$ and $R^{(1)}_{+} - R^{(1)}_{-} +  {\widetilde R}^{(2)}_{-} - {\widetilde R}^{(2)}_{+}>0$ then $J^{(1)}_+$ increases while $J^{(1)}_-$ decreases until $J^{(1)}_+=1$ and $J^{(1)}_- =0$ so that the measure of $\vert\Phi^{(1)}\rangle$ gives  $\vert\phi^{(1)}_{+}\rangle $. By the same token, 
and the same condition $R^{(1)}_{+} - R^{(1)}_{-} +  {\widetilde R}^{(2)}_{-} - {\widetilde R}^{(2)}_{+}>0$, if $J^{(2)}_{+}\neq 0$ then $J^{(2)}_-$ increases until $J^{(2)}_-=1$ and $J^{(2)}_+ =0$ so that the measure of $\vert\Phi^{(2)}\rangle$ gives  $\vert\phi^{(2)}_{-}\rangle $. And this is the necessarily expected result.

The final outcome of the measure is determined by the evolution of $\vert\Phi^{(1)}\rangle \vert\Phi^{(2)}\rangle$ which depends on the values that the fields ${\mathscr X}_\pm({\bf x}, t_0,\tau_0+\Delta\tau)$ had immediately before the measurement at ${\bf x}_1$ and ${\bf x}_2$, and thus it depends on the random component of the field and on the unknown details about the shaping of the information-carrying impulses. All this makes the individual outcomes of measurements on entangled entities unpredictable, outcomes which however remain perfectly correlated.
In other words, although it is demonstrated that the Bell inequality is violated by non-product states \cite{gisin}, nevertheless, also with the assumption done in the present work  that a Bell state factorizes when both entangled entities interact simultaneously with the respective measurement devices, the physical information about the entanglement - and thus the violation of Bell inequality - is maintained by the collapse equations  entailing correlated outcomes of the measures via the field ${\mathscr X}({\bf x}, t,\tau)$.
\begin{figure}[h!]
 \centering
 \includegraphics[scale=0.35,keepaspectratio=true,angle=0]{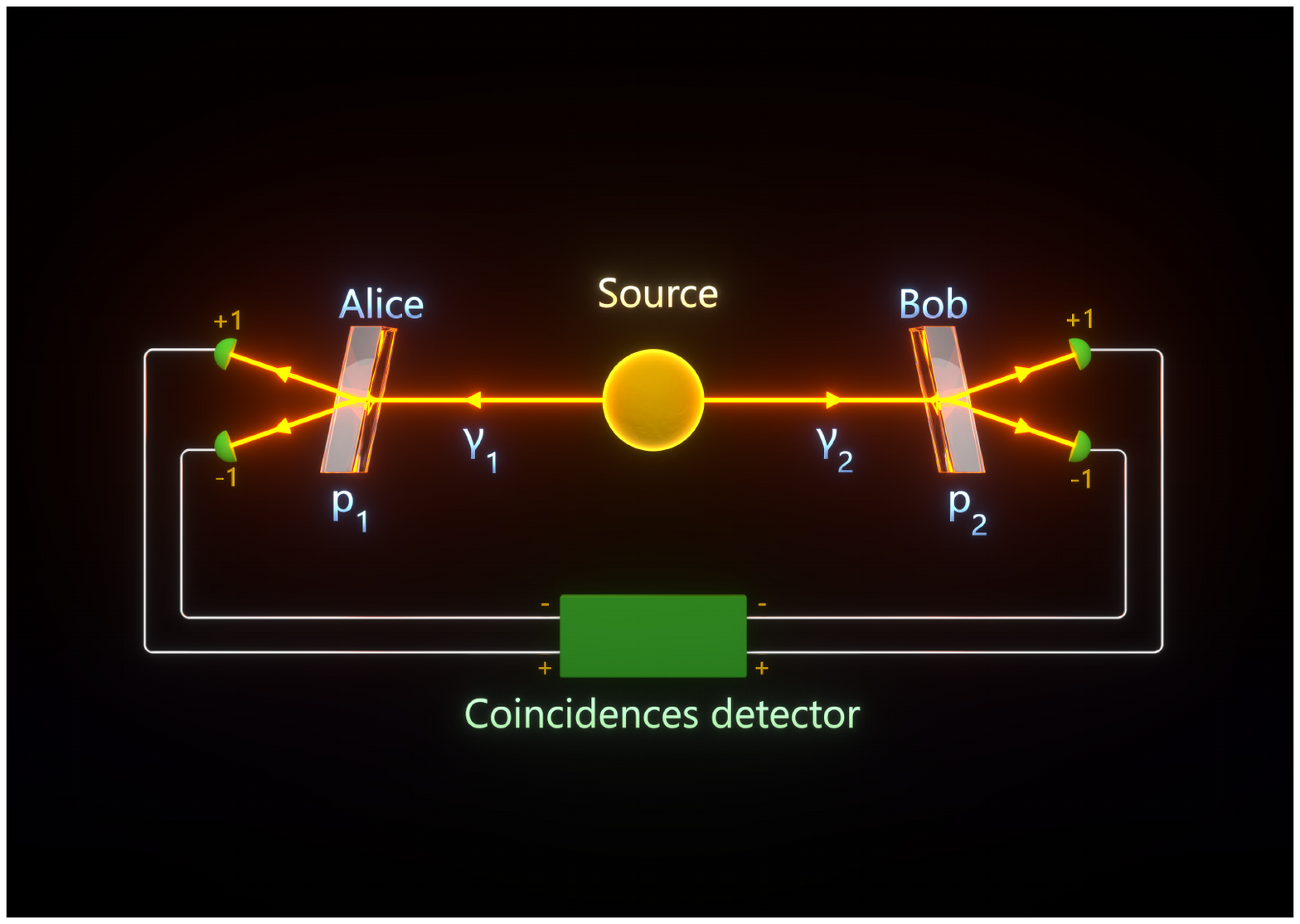}
 \caption{\sl{An apparatus for performing a Bell test via quantum correlation measurement. A source $S$ emits a pair of entangled photons $\gamma_1$ and $\gamma_2$ in a singlet state $\vert\Psi\rangle$ and their linear polarizations are measured by polarizers $p_1$ and $p_2$. Each polarizer has two output channels, labeled +1 and -1. In this schematic diagram a representation of the switching devices, performing a change of the settings of the polarizers while the photons are in flight between the source and the polarizers, is omitted.}  }
\label{fig2}
\end{figure}
\subsection{Proposal for a nonconventional experiment} \label{experiment}
Quantum mechanics predicts random results on each side of an experimental apparatus like the one sketched in Figure \ref{fig2} with $1/2$ probability of measuring $+1$ or $-1$, and it also predicts strong correlations between these random results.  Bell’s inequality gives an upper bound to the correlations predicted by local realism, whereas quantum predictions violate this inequality. A Bell test consists of measuring the correlations and comparing the results with Bell’s inequality. To perform a loophole-free  Bell test, the polarizer settings must be changed randomly while the photons are in flight between the source and the polarizers \cite{aspect1,aspect2,aspect3}.
\begin{figure}[h!]
 \centering
 \includegraphics[scale=0.35,keepaspectratio=true,angle=-90]{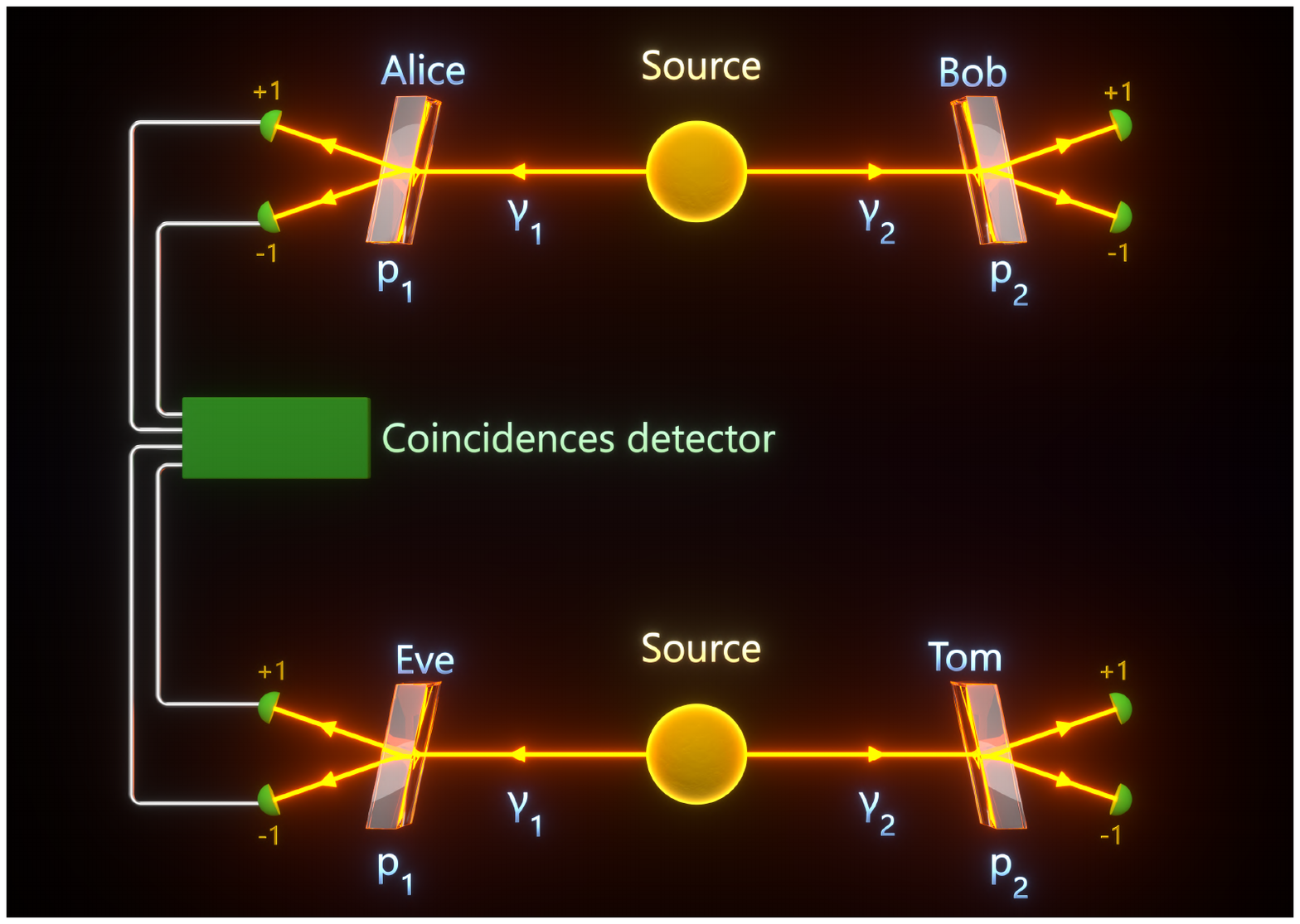}
 \caption{ \sl{Two identical and standard apparatuses  - used to check Bell's inequality violation in coincidences measurement of entangled particles - should be operated at the same time. The two sources emit entangled photons $\gamma_1$ and $\gamma_2$ whose polarizations are analysed by the polarizers $p_1$ and $p_2$. Each polarizer has two possible outputs indicated as $+1$ or $-1$. The spatial arrangement should be such that the distance between Alice and Eve is much shorter than that between Alice and Bob and Eve and Tom, respectively. The two setups should be suitably electromagnetically shielded in order to avoid loopholes. A coincidences detector should be used to check whether or not Bell's inequality is violated when checking the correlations between the measurements performed by Alice and Eve. In case of a positive result this would lend credit to the idea surmised in the present work.
 (The experiment could as well be performed using particles with spin)}.}
\label{fig3}
\end{figure}
The source term ${\widetilde{F}}[{\bf x}, \psi_{\pm}^{(j)}(t), \tau ; t_0, \tau_0]$ in equation \eqref{eqperxi}, even if of unspecified analytic form, amounts to assuming that the field  ${\mathscr X}({\bf x}, t, \tau)$ carries the information about the wavefunction collapse due to the measurement performed at some  given place, and equation \eqref{eqperxi} tells that this information is spread in every spatial directions \color{black} through the extra time dimension 
 $\tau$. Let us repeat once more that entanglement cannot be explained through an instantaneous information transfer in our (3,1) space-time.\color{black}
Then equations \eqref{BBextended} describe the entanglement between two objects as due to their correlated collapses of the respective wavefunctions   driven by a mutual transmission of information by means of the field ${\mathscr X}({\bf x}, t, \tau)$  through the extra time $\tau$. Therefore, the collapse information of a given system could also reach another identical system driving its wavefunction collapse, so mimicking entanglement correlation, even if these two systems are not entangled according to the quantum formalism \cite{nota2}. Hence the violation of Bell inequality could be unexpectedly  found also in apparently trivial situations.\color{black}
This is the hypothesis to be experimentally proved or disproved that also suggests how to design an appropriate experiment. 
\color{black}
An experiment that would appear meaningless if compared to the standard ways of considering the violation of Bell inequality reported in the literature.
To the contrary, it will turn out meaningful if the working hypothesis advanced in the present paper has a counterpart in the physical world.
In fact,
let us now consider two identical EPR sources, schematically depicted in Figure \ref{fig3}, simultaneously emitting each a pair of entangled photons or particles. \color{black} An experimental method for simultaneous generation of two independent pairs of entangled photons for example is reported in Ref.\cite{zeilinger}. \color{black}
Alice and Bob perform measurements at the ends of the first system, and Eve and Tom at the ends of the second system. 
Of course, the photons (or particles with spin) reaching Alice and Bob or Eve and Tom are in state superposition and constrained to fulfil a conservation
law (momentum, energy, angular momentum) which is not the case of two particles reaching, say, Alice and Eve because a-priori the entities (say photons, as in Figure \ref{fig3}) reaching Alice and Eve are independent, that is, not entangled. 
 Now, as soon as the photon detected by Alice has "chosen" its polarization state how does the photon being detected by Bob "know" that it has no choice  left for its polarization state?
The orthodox reply is that they are part of the same indivisible system even if they are very far apart, but if we assume that the information about the outcome of the measurement performed by Alice is broadcasted in every spatial direction (through the extra temporal dimension), a particle belonging to  an identical system, like the one reaching Eve, could be driven 
to collapse into the same complementary state as in Bob's case.   
 
We could wonder why should we assume that Eve’s particle reacts to the information broadcast from Alice’s particle, or vice versa, and why it would not be more natural to assume that Alice’s and Bob’s particles only respond to each other’s information exchange, and similarly for the pair Eve and Tom. 
After the basic assumption that Alice's and Bob's particles respond to each other's information exchange via the field ${\mathscr X}({\bf x}, t, \tau)$, which is transmitted by each particle when hitting its measuring device, when we move on to consider four identical measuring devices receiving identical particles emitted synchronously from identical sources, in the absence of any hypothesis about how the field ${\mathscr X}({\bf x}, t, \tau)$ might encode from which particular observer the information about the collapse of the wave function comes, the simplest assumption is that Alice's and Eve's particles can also respond to each other's information exchange via the field ${\mathscr X}({\bf x}, t, \tau)$ under suitable conditions as follows. \color{black} So, let's imagine an experimental setup in which Alice's and Eve's measuring apparatuses are spatially close to each other and both much further away from the apparatuses of their companions Bob and Tom,
so that the extra-time interval $\Delta_1\tau$ needed by the hypothetical field ${\mathscr X}({\bf x}, t, \tau)$ to exchange information between Alice and Bob on one side, and Eve and Tom on the other side, is much longer than the interval $\Delta_2\tau$ needed to exchange information between Alice and Eve. 
Now, the global state vector of the two independent subsystems - Alice, Bob and Eve, Tom - is the product 
$\vert\Psi_{A,B}\rangle \vert\Psi_{E,T}\rangle $ 
where
\begin{eqnarray}\label{BellStateABET}
\vert\Psi_{A,B}\rangle  &=&\frac{1}{\sqrt{2}}(\vert\phi^{(A)}_{+}\rangle\vert\phi^{(B)}_{-}\rangle - \vert\phi^{(A)}_{-}\rangle\vert\phi^{(B)}_{+}\rangle )\\
\vert\Psi_{E,T}\rangle  &=&\frac{1}{\sqrt{2}}(\vert\phi^{(E)}_{+}\rangle\vert\phi^{(T)}_{-}\rangle - \vert\phi^{(E)}_{-}\rangle\vert\phi^{(T)}_{+}\rangle )\  .
\end{eqnarray}
If the four measuring devices in the setup of Fig.\ref{fig3} simultaneously detect the photons, or particles, then the state vector $\vert\Psi_{A,B}\rangle \vert\Psi_{E,T}\rangle$ factorizes to $\vert\Phi^{(A)}\rangle \vert\Phi^{(B)}\rangle\vert\Phi^{(E)}\rangle \vert\Phi^{(T)}\rangle$ [with the notations of \eqref{Phi1})] out of which - after the above given argument and experimental setup - correlated behaviours are expected within the state 
$\vert\Phi^{(A)}\rangle \vert\Phi^{(E)}\rangle$. 
Consequently, the central hypothesis of the present work boils down to rewriting the collapse equations \eqref{BBextended} as 
\begin{eqnarray}\label{AliceEve}
\frac{dJ^{(A)}_{+}}{dt} &=&2 \gamma (t) \left( R^{(A)}_{+} - R^{(A)}_{-} + {\widetilde R}^{(E)}_{-} - {\widetilde R}^{(E)}_{+}\right)J^{(A)}_{+} J^{(A)}_{-}  \nonumber\\
\frac{dJ^{(A)}_{-}}{dt} &=& 2\gamma (t)\left( R^{(A)}_{-} - R^{(A)}_{+}  + {\widetilde R}^{(E)}_{+} - {\widetilde R}^{(E)}_{-} \right)J^{(A)}_{-} J^{(A)}_{+} \\
\frac{dJ^{(E)}_+}{dt} &= &2\gamma (t)\left( R^{(E)}_{+} - R^{(E)}_{-} + {\widetilde R}^{(A)}_{-} - {\widetilde R}^{(A)}_{+}\right)J^{(E)}_{+} J^{(E)}_{-}  \nonumber\\
\frac{dJ^{(E)}_-}{dt} &=& 2\gamma (t)\left( R^{(E)}_{-} - R^{(E)}_{+}  + {\widetilde R}^{(A)}_{+} - {\widetilde R}^{(A)}_{-} \right)J^{(E)}_{-} J^{(E)}_{+}  \nonumber
\end{eqnarray}
where
\begin{equation}
 R^{(A,E)}_{\pm} = \frac{\vert\psi^{(A,E)}_{\pm}(t)\vert^2}{\vert{\mathscr X}_\pm({\bf x}_{A,E}, t_0, \tau_0+\Delta_2\tau  )  \vert^2}
\end{equation}
\begin{equation}
{\widetilde R}^{(A)}_{\pm} = \frac{\vert\psi^{(A)}_{\pm}(t)\vert^2}{\vert{\mathscr X}_\pm({\bf x}_E - \textbf{w}\Delta_2\tau , t_0, \tau_0+\Delta_2\tau )  \vert^2} \ ,\quad\quad\quad 
{\widetilde R}^{(E)}_{\pm} = \frac{\vert\psi^{(E)}_{\pm}(t)\vert^2}{\vert{\mathscr X}_\pm({\bf x}_A - \textbf{w}\Delta_2\tau ,  t_0, \tau_0+\Delta_2\tau )  \vert^2}
\end{equation}
with $\Delta_2\tau = \vert{\bf x}_A - {\bf x}_E\vert/\vert\textbf{w}\vert$.
If the equations \eqref{AliceEve} describe a real physical process or are meaningless is ascertained by performing a coincidence measurement between Alice and Eve instead of between Alice and Bob or Eve and Tom.
Thus, denote with $P_{\pm\pm}(\alpha, \beta)$ the probabilities of obtaining the local results $\pm 1$ in the direction $\alpha$ for the particle detected by Alice and $\pm 1$ in the direction $\beta$ for the particle detected by Eve, $\alpha$ and $\beta$ being the directions of the polarization analyzers (e.g. identified via the angle with respect to a reference direction), the correlation coefficient given by
\[
E(\alpha,\beta) = P_{++}(\alpha, \beta) + P_{--}(\alpha, \beta) - P_{+-}(\alpha, \beta) - P_{-+}(\alpha, \beta)
\]
enters the Bell-CHSH (Clauser, Horne, Shimony, Holt) \cite{clauser,aspect2} inequality 
\begin{equation}
S(\alpha, \alpha^\prime,\beta,\beta^\prime) = [E(\alpha,\beta) -  E(\alpha,\beta^\prime)] +  [E(\alpha^\prime,\beta) +  E(\alpha^\prime,\beta^\prime)]\le 2
\end{equation}
which is satisfied by realistic local theories, and a-priori should be also fulfilled by non entangled particles as those reaching Alice and Eve.
Quantum mechanics for various combinations of directions of polarization analyzers $\alpha, \alpha^\prime,\beta,\beta^\prime$ predicts a violation of this inequality, the violation being maximal for the set of angles $S_{QM}(0, 45^{o}, 22.5^{o},67.5^{o}) = 2 \sqrt{2}$. Therefore, let us imagine to measure this quantity through the coincidences detected between Alice and Eve - under all the standard requirements of detectors efficiency, electromagnetic shielding to avoid loopholes and so on - in case one would observe a violation of the above Bell-CHSH inequality this would support the hypothesis of the existence of an information exchange - between the subsystems - mediated by some physical entity that we have called sub-quantum field  
${\mathscr X}({\bf x}, t, \tau)$ propagating through an extra time dimension. This hypothesis is thus falsifiable and even if somewhat daring it cannot be a-priori discarded.

A clarification needs to be made here.  Equations \eqref{BBextended} and \eqref{AliceEve} describe the evolution of wavefunctions collapse in ordinary time after $t_0$, the instant of particles interaction with the measuring devices, and "after" a mutual exchange of information in the extra time. In the case of the proposed new experiment, the surmised phenomenon of information exchange through the extra time depends entirely on the spatial arrangement of the four measuring devices. The  exchange of information between Eve and Alice (assumed at the shortest distance among the four measuring devices) could conceivably destroy the entanglement with their respective partners allowing a rearrangement of the correlations among the results of the four measuring devices.
A more refined analysis would be needed to describe what kind of correlations - among all these devices - could be expected with generic spatial arrangements. Such analysis is at present beyond the scope of our toy model.

An important remark is in order. What is described in this Section aims to propose the conceptual scheme of a possible experimental verification of the physical hypothesis  underlying this work on quantum entanglement. Indeed, the practical implementation of this proposal could make use of  recent  developments to design a more sophisticated experimental setup than the one sketched in Fig.\ref{fig3}, resorting for example to the use of solid-state devices where the phenomenon of parametric down-conversion occurs inside a non-linear crystal generating entangled photons \cite{physrepGenovese,mandel,kwiat1}, as well as various methods to ensure loophole-free experiments \cite{physrepGenovese,kwiat2}.  

\section{Concluding remarks} \label{conclusions} 

Entanglement, that is the ability of two particles to "correlate" their behaviors even at very large distances and in the absence of any physical connection, is intrinsic to the formalism of quantum mechanics and although it is an experimentally proven reality of our world, it defies our perception of physical phenomena  and the way of representing them. On the other hand, since its inception, quantum mechanics has always allowed to perform very accurate calculations regarding microscopic phenomena, providing predictions that have always been verified without exception. So why should we worry about 
entanglement since the theory works so well? But attempting to delve behind a formalism describing as an indivisible system two entangled objects,  even if sitting at opposite borders of our galaxy, could have a twofold interest, on the one side a conceptual relevance and, on the other side, perhaps interesting implications in the field of quantum technologies and quantum computation where entanglement plays a crucial role.
Therefore, any attempt at gaining a deeper understanding of quantum entanglement through different hypotheses seems worthwhile, even in case a given hypothesis is disproved by experiments, since it would thus introduce a "no go".
A possible "softening" of the conundrum represented by quantum entanglement could be found by considering what is observed in our 3+1 dimensional space-time as a projection from a higher dimensional space-time.
This has been suggested until recently \cite{genovese} by invoking the existence of extra dimensions of space kind, in fact, two points very far apart in the 3+1 space-time can be very close one another in a higher dimensional embedding space.
In the present work we have suggested a different scenario by invoking the existence of an extra dimension of time kind in order to describe  \textit{nonlocality without nonlocality}, that is, nonlocal correlations without nonlocal causation, in the words of the author of Ref.\cite{weinstein}.
Aiming at depicting a possible scenario of this kind, we have borrowed the longstanding Bohm-Bub's phenomenological proposal for a non-unitary dynamical description of the wavefunction collapse \cite{BB,tutsch}, and, after suitable elaboration, we have outlined a toy model whose function is to motivate an experiment to test a physical hypothesis: the existence of a sub-quantum field carrying information. An experimental test that would have no reason to be performed, being manifestly trivial in the absence of any thinkable reason to doubt of its triviality. Of course, the model put forward in the present work can be criticized from many different viewpoints but, being not conflicting with the robust theoretical framework of quantum mechanics, it is intended to provide the mentioned thinkable reason to stimulate the interest of some experimentalists, and, possibly, some constructive theoretical contribution.  In fact, the proposed experiment is clearly defined and not affected by the unspecified details of the phenomenological toy model. 

Moreover, it is worth mentioning that the puzzling and counterintuitive properties of entanglement between spacelike separated quantum objects turn to unbelievable when entanglement is found
between photons that never coexisted in time \cite{entanglTime}. Out of two temporally separated photon pairs it is found that one photon belonging to the first pair can be entangled with a photon from the second pair and the first photon is detected before the creation of the second one.
This experimental outcome followed a previous theoretical work where quantum interferences and violation of Bell’s inequality has been studied by considering photons emitted from independent single photon sources that do not overlap in time\cite{zanthier}. At least in principle, we can speculate that this phenomenon could be given a more "intuitive" explanation under the hypothesis put forward in the present paper by considering that the  information carrying field ${\mathscr X}({\bf x}, t, \tau)$ a-priori should propagate also in our familiar time coordinate $t$. This is not explicitly formalized in equation \eqref{evolEqForXi} because the measuring process was assumed to take place during a very short interval of ordinary time.
In fact, equation \eqref{evolEqForXi} could be expressed as a ultrahyperbolic wave equation thus propagating to the ordinary future the information of the outcome of the measure on the first photon. This information could thus drive the outcome of the measure on the second photon.  
 
Finally, in case the proposed experiment would lend credit to the hypothesis formulated in the present work,  information would be given an ontological status and in so doing this would be somehow echoing \textit{"[...] the idea that every item of the physical world has at bottom — at a very deep bottom, in most instances — an immaterial source and explanation; [...] in short, that all things physical are information-theoretic in origin..."} as advocated by J.A. Wheeler \cite{wheeler}.

\bigskip

\section{appendix}
Let us sketchily show how we can borrow from Ref.\cite{visser} - almost \textit{verbatim} - the suggestion for a five dimensional space-time metric of (+ + + - -) signature - with a non-compact and infinite extra time dimension -  where the particles are trapped on the standard four-dimensional space-time. Thus, by simply modifying the signature of the extra dimension of the "exotic" Kaluza-Klein metric of 
Ref.\cite{visser}, consider the infinitesimal arc-length in 3+2 space-time as 
$$ds^2 = g_{\mu\nu} dx^\mu dx^\nu =  dx^2 + dy^2 + dz^2 - e^{2\varphi(\tau)}c^2 dt^2 - w^2 d\tau^2 $$
then the Klein-Gordon equation written on this space-time background (using the Laplace-Beltrami operator) reads
\begin{equation}
g^{-1/2} \partial_\mu(g^{1/2}g^{\mu\nu}\partial_\nu\Psi) - m_5^2\Psi = 0
\end{equation}
where $g$ is the determinant of the metric $g_{\mu\nu}$ and $m_5$ is a suitably defined five dimensional particle rest mass \cite{visser}. One finds
\begin{equation}
\left[ \nabla^2 - e^{- 2\varphi(\tau)}\frac{1}{c^2}\frac{\partial^2}{\partial t^2} - \frac{1}{w^2}\frac{\partial^2}{\partial\tau^2} \right]\Psi - 
\frac{1}{w^2}\frac{\partial\varphi}{\partial\tau}\frac{\partial\Psi}{\partial\tau} - m_5^2\Psi = 0
\end{equation}
which has the form a ultrahyperbolic equation \cite{weinstein} for which a solution can be found in the form of travelling waves in the four standard directions but confined in the extra time direction $\tau$, that is 
\begin{equation}
\Psi = e^{ -i (\omega t - \bf{k\cdot x})} e^{-\varphi(\tau)/2} \Phi(\tau)
\end{equation}
where $\Phi(\tau)$ satisfies the equation
\begin{equation}
\left[\frac{1}{w^2}\frac{\partial^2}{\partial\tau^2}+ (\frac{1}{2} \ddot\varphi + \frac{1}{4}\dot\varphi^2 -\omega^2e^{-2\varphi}) + m_5^2 + {\bf{k}}^2\right]\Phi =0
\end{equation}
then with a convenient choice of $\varphi(\tau)$,  the factor $e^{-\varphi(\tau)/2}$ - which is called warp factor -  determines the degree of warping along the extra time dimension. 
These are just a few hints on how to imagine an infinite temporal extra-dimension where the sub-quantum field ${\mathscr X}({\bf{x}}, t, \tau)$ is fully extended whereas particles are trapped around $\tau =0$. The field ${\mathscr X}({\bf{x}}, t, \tau)$ is assumed to be a real physical entity, therefore in principle it could enter the $3+2$ space-time metric $g_{\mu\nu}$ similarly to the electromagnetic vector potential in Kaluza-Klein theories. This could be done so as to find a field equation for ${\mathscr X}({\bf{x}}, t, \tau)$ as that in Eq.\eqref{eqperxi}. 
However, how to choose such a metric and how to choose a suitable and physically meaningful function 
$\varphi(\tau)$  remain far beyond the aim of the present work.

\section*{Acknowledgments}
The author wishes to thank Roger Penrose for an interesting and useful discussion held at the Arts Centre De Brakke Grond, Amsterdam, in 2014, a discussion that has been the remote origin of the present work. Useful comments and suggestions emerged during several  discussions with Giulio Pettini, Gabriele Vezzosi, Matteo Gori, Roberto Franzosi, Guglielmo Iacomelli, and Jack Tuszynski. The author thanks Stefano Ruffo for having brought to his attention the paper in Ref.\cite{baracca}. This work was partially supported by the European Union’s Horizon 2020 Research and Innovation Programme under Grant Agreement No. 964203 (FET-Open LINkS project). {{The author also thanks two anonymous referees for their useful comments that helped to improve the clarity of the presentation of this work}}.




\begin{thebibliography}{99}
\bibitem{nota} There are experimental results about the violation of Bell inequality also at large distances (see for instance  \protect\cite{nl2,nlz}) however there is no available information about a possible change of the degree of violation of Bell inequality as a function of the distance between entangled entities.
\bibitem{nota2} This means that the state vector describing the ensemble of the two systems is the product of the state vectors of the two systems.
\end{thebibliography}
\end{document}